\documentclass[12pt,oneside]{article}
\usepackage{amsmath, amsbsy, amsthm, amsfonts, array,graphics}
\usepackage{graphicx}
\usepackage[round]{natbib}
 \DeclareMathOperator{\sgn}{sgn}
 \DeclareMathOperator{\der}{d}
 \DeclareMathOperator{\E}{E}
 \DeclareMathOperator{\Prob}{pr}

 \newcommand{\ipara}{\psi} %para of interest
 
 \newcommand{\hipara}{\hat{\psi}}
 \newcommand{\iparanull}{\psi_0}
 \newcommand{\npara}{\lambda} %nusiance para
 \newcommand{\Npara}{\Lambda}
 
 \newcommand{\bnpara}{\lambda}
 \newcommand{\hbnpara}{\hat{\lambda}}
 \newcommand{\para}{\omega}
 \newcommand{\bpara}{\omega}
 
 \newcommand{\hbpara}{\hat{\omega}}

\title{\textbf{ \large{A practical procedure to find matching priors for
frequentist inference}}}
\author
{\normalsize{\textsc{By JUAN ZHANG and JOHN E. KOLASSA}}\\
\it{\footnotesize{Department of Statistics,
Rutgers University,}}\\
\it{\footnotesize{Piscataway, New Jersey, 08854, U.S.A.}}\\
\footnotesize{janezh@stat.rutgers.edu} \ \ \footnotesize{
kolassa@stat.rutgers.edu} }

\date{}

\begin{document}
\maketitle

\begin{center}
    \textsc{Summary}
\end{center}

We present a practical way to find the matching priors proposed by
\citet{WP} and \citet{Pe}. We investigate the use of saddlepoint
approximations combined with matching priors and obtain $p$-values
of the test of an interest parameter in the presence of nuisance
parameter. The advantage of our procedure is the flexibility of
choosing different initial conditions so that one can adjust the
performance of the test. Two examples have been studied, with
coverage verified via Monte Carlo simulation. One relates to the
ratio of two exponential means, and the other relates the logistic
regression model. Particularly, we are interested in small sample
settings.

\vskip 0.5cm \noindent
\begin{footnotesize}{\textit{Some key words:} Bayes; Conditional
inference; Matching prior; Modified signed root likelihood ratio
statistic; Partial differential equation; Saddlepoint
approximation.}\end{footnotesize}

\newpage

\begin{center}
    \textsc{1. Introduction}
\end{center}

We consider inference on a single scalar parameter in the presence
of nuisance parameters.   Under the frequentist settings,
conditional inference can be complicated. Bayesian method can
simplify frequentist elimination of nuisance parameters.  The
frequentist and the Bayesian approaches can be connected by matching
priors. Matching priors were first proposed by \citet{WP} and
\citet{Pe}. Determining a matching prior is equivalent to finding a
solution of a first order partial differential equation. Only in
simple circumstances, such as when parameters are orthogonal, the
partial differential equation can be solved analytically. \citet{lc}
note that ``Unfortunately, except for these cases, the solution of
the resulting partial differential equations becomes quite a hurdle;
our only hope is to find numerical solutions to these partial
differential equation."

We will see a practical way to solve for the matching priors,
without the involvement of the back transformation described by
\citet{lc}. This procedure is easy to understand, can be implemented
in R, \citet{R} and is suitable to all kinds of initial conditions.

Our implementation of matching priors for the approximations
proposed by \citet{DM} is less complicated than other frequentist
methods. DiCiccio and Martin's approximations are saddlepoint
approximations that make use of Bayesian--frequentist parallels. Our
proposed implementation requires less computational effort compared
to the iterative Metropolis-Hasting algorithm described by
\citet{lc}.

We end the introduction with a brief outline of this paper.  In \S
2, We review the concepts of matching priors and discuss the
circumstance when orthogonal parameters are presence. Existing
analytical and numerical solutions are reviewed. In \S 3, we present
the procedure for solving matching priors, both analytically and
numerically. Specification of initial condition is discussed. We
also provide information of R software implementation of the solving
procedure. In \S 4, the approximations of \citet{DM} are reviewed.
The application of using matching priors conjuncted with DiCiccio
and Martin's approximations is illustrated through examples in
section \S 5. Different initial conditions are specified for
obtaining various matching priors. Finally, \S 6 contains the
conclusion.

\vskip 0.8cm
\begin{center}
    \textsc{2. Matching priors}
\end{center}

We consider parametric models with random variables $X_1,\dots,X_n$
having joint density function that depends on the unknown parameter
vector $\bpara$.  Suppose $\bpara$ is of length $d$ and
$\bpara=(\para^1,\para^2,\dots,\para^d)=(\ipara,\bnpara)$ with
$\ipara=\para^1$, the parameter of interest, and the nuisance
parameter $\bnpara=(\para^2,\dots,\para^d)$.

Matching priors were proposed by \citet{WP} and \citet{Pe}. In the
following, denote the matching prior by $\pi(\cdot)$. Let
$\Prob_{\pi}(\cdot|X)$ be the posterior probability measure for
$\ipara$ under prior $\pi(\cdot)$.  The upper $(1-\alpha)$ posterior
quantile constructed on the basis of a prior density function
$\pi(\ipara)$ has the property that it is also the frequentist
limit, such that
    $$\Prob_{\pi} \{\ipara\le\ipara^{(1-\alpha)}(\pi,X)|X\}=
    \Prob_{\ipara}\{\ipara\le\ipara^{(1-\alpha)}(\pi,X)\}
    =1-\alpha+O(n^{-1}).$$

When there are no nuisance parameters, \citet{WP} showed that the
appropriate choice of $\pi(\para)$ is
    $\pi(\para)\propto\{i(\para)\}^{1/2},$
where $i(\para)=\E \{-\der ^2  l (\para)/\der \para^2\}$, and
$l(\cdot)$ is the log-likelihood function.  In this case, matching
priors can be easily obtained.

In the presence of nuisance parameters, \citet{Pe} showed that
$\pi(\bpara)$ must be chosen to satisfy the partial differential
equation
    \begin{equation} \label{prior}
        \sum_{j=1}^d i^{1j} (i^{11})^{-1/2}
        \frac{\partial}{\partial \para^j} (\log \pi)
        +\sum_{j=1}^d \frac{\partial}{\partial \para^j}
        \{i^{1j}(i^{11})^{-1/2}\}=0,
    \end{equation}
where $i_{jk}(\bpara)=\E \{-\partial^2
l(\bpara)/\partial\para^j\partial\para^k\}$ and $(i^{jk})$ is the
$d\times d$ inverse matrix of $(i_{jk})$.

If the parameter of interest and the nuisance parameter vector are
orthogonal, solving the partial differential equation (\ref{prior})
is relatively easy.  We follow the definition of parameter
orthogonality by \citet{cr87}. Orthogonality is defined with respect
to the expected Fisher information matrix. The most direct
statistical interpretation of parameter orthogonality is that the
relevant components of the original statistic are uncorrelated. In
general, it is possible to obtain orthogonality of a scalar
parameter of interest to a set of nuisance parameters.

When the parameter of interest $\ipara$ is orthogonal to a set of
nuisance parameters, equation (\ref{prior}) reduces to
    \begin{align}\label{othprior}
        (i_{\ipara\ipara})^{-1/2}
        \dfrac{\partial}{\partial \ipara}(\log \pi)
        +\dfrac{\partial}{\partial
        \ipara}(i_{\ipara\ipara})^{-1/2}=0.
    \end{align}
\citet{tib89} showed that solutions were of the form
$\pi(\ipara,\bnpara)\propto\{i_{\ipara\ipara}(\ipara,\bnpara)\}^{1/2}$
$g(\bnpara)$, where $g(\bnpara)$ is arbitrary, and the suggestive
notation $i_{\ipara\ipara}(\ipara,\bnpara)$ is used in place of
$i_{11}(\ipara,\bnpara)$.

However, choosing a parametrization to achieve parameter
orthogonality is not always easy, and it can be hard in some cases.
It is equivalently hard to obtain orthogonalization and to solve the
partial differential equation (\ref{prior}) directly, since the
orthogonalization procedure also requires solutions to partial
differential equations of form similar to (\ref{prior}).  Staicu and
Reid (2007), studied the use of matching priors with the
approximation of \citet{DM} under orthogonal parametrization, and
showed that the Peers-Tibshirani class of matching priors was
essentially unique. One can modify the arguments in this paper to
solve the partial differential equation that defines the
orthogonality transformation, and attempt, using orthogonality, to
narrow down the class of matching priors.

\citet{lc} proposed a general procedure to solve the partial
differential equation (\ref{prior}) numerically, in models with a
single nuisance parameter.  Firstly, they transform the parameters
into another parameter space, solving the equation, and then
transform back to the original parameter space.  The numerical
application of this procedure is not necessarily easy, and usually
the transformation between the two parameter spaces is nontrivial.
\citet{lc} implemented their procedure using Mathematica.  They did
not give instructions on initial condition specification, which is a
necessary component to give specific solution in solving the partial
differential equation. \citet{s} introduced data-dependent priors
that locally approximate the matching priors, and his procedure can
deal with vector nuisance parameters.

\vskip 0.8cm

\begin{center}
    \textsc{3. Solving for the matching priors}
\end{center}

In this section, we introduce a procedure to solve the partial
differential equation (\ref{prior}) in general parametrization.  For
simplicity, we consider the dimension of the parameter space as 2.
First, we give analytical form of the solutions, and then practical
notes will be presented later in this section.

In the case that $d=2$, equation (\ref{prior}) is reduced to
    \begin{equation}\label{pde}
        a(\ipara,\npara)z_{\ipara}+b(\ipara,\npara)z_{\npara}
                     =d(\ipara,\npara),
    \end{equation}
where $$z(\ipara,\npara)=\log \{\pi(\ipara,\npara)\},$$
$$a(\ipara,\npara)=\{i^{11}(\ipara,\npara)\}^{1/2},$$
$$b(\ipara,\npara)=i^{12}(\ipara,\npara)\{i^{11}(\ipara,\npara)\}^{-1/2},$$
and
    $$d(\ipara,\npara)=-\left[\frac{\partial}{\partial
        \ipara}\{i^{11}(\ipara,\npara)\}^{1/2}+\frac{\partial}{\partial
        \npara}\{i^{12}(\ipara,\npara)\}\{i^{11}(\ipara,\npara)\}^{-1/2}\right].$$
The coefficient $a(\ipara,\npara)$ is a diagonal element of the
inverse matrix of $(i_{jk})$, so $a(\ipara,\npara)$ can not be zero.
Dividing both sides of (\ref{pde}) by $a(\ipara,\npara)$, we have
    \begin{equation*}\label{pdeuse}
        z_{\ipara}+\frac{b(\ipara,\npara)}{a(\ipara,\npara)}z_{\npara}
            =\frac{d(\ipara,\npara)}{a(\ipara,\npara)}.
    \end{equation*}
This forces the coefficient of $z_{\ipara}$ to be 1, which
simplifies the procedure of finding a solution.

To solve the equation (\ref{prior}), it suffices to solve the
following ordinary differential equations system
    \begin{align}\label{odes}
    \dfrac{\der \ipara}{\der s}=1,\dfrac{\der \npara}{\der s}
            =\dfrac{b(\ipara,\npara)}{a(\ipara,\npara)},
        \dfrac{\der z}{\der s}
            =&\dfrac{d(\ipara,\npara)}{a(\ipara,\npara)}.
    \end{align}

To be more specific with the solution, let us consider the initial
conditions prescribed along an initial curve $I$.  Suppose that $I$
is given parametrically, in terms of a parameter $\xi$, as
    \begin{equation*}
        \ipara=\Psi(\xi),\ \ \npara=\Npara(\xi).
    \end{equation*}
Then evaluating $z(\ipara,\npara)$ at a point on $I$ is equivalent
to expressing $z$ as a function of $\xi$,
    \begin{equation}\label{uini}
        z=Z(\xi)=z\{\Psi(\xi),\Npara(\xi)\}.
    \end{equation}
Here, it is obvious to see that $I$ can not be tangent to the
direction
$\left[1,\dfrac{b\{\Psi(\xi),\Npara(\xi)\}}{a\{\Psi(\xi),\Npara(\xi)\}}\right]$.
We then obtain
    \begin{equation*}\label{odesol}
        \ipara=\ipara(s,\xi),\ \ \npara=\npara(s,\xi)
    \end{equation*}
by simultaneously integrating the two equations defined by
    \begin{align}
        \label{ode1} \frac{\der \ipara}{\der s}=&1,
                \ \ \ \ipara(s_0,\xi)=\Psi(\xi),\\
        \label{ode2} \frac{\der \npara}{\der s}
                =&\frac{b(\ipara,\npara)}{a(\ipara,\npara)},\ \ \
                        \npara(s_0,\xi)=\Npara(\xi).
    \end{align}
From the third equation in (\ref{odes}), the initial condition is
given by (\ref{uini}).  Then we have,
    \begin{equation}\label{ode3}
        \frac{\der z}{\der s}=\frac{d(\ipara,\npara)}{a(\ipara,\npara)},
                \ \ \ z(s_0,\xi)=Z(\xi),
    \end{equation}
Equation (\ref{ode3}) can be integrated by quadrature, once
equations (\ref{ode1}) and (\ref{ode2}) have been solved,
    \begin{equation}\label{zsol}
        z(s,\xi)=Z(\xi)+\int^s_{s_0}\frac{d\{\ipara(s',\xi),\npara(s',\xi)\}}
            {a\{\ipara(s',\xi),\npara(s',\xi)\}}\der s'.
    \end{equation}
These generate a surface in three dimensions, $Z(\ipara,\npara)$,
that satisfies both the equation (\ref{pde}) and the initial
condition. When there are no close form solutions for equations
(\ref{ode1}),(\ref{ode2}) and (\ref{ode3}), numerical solutions can
be achieved.  \citet{PDE} presents more mathematical details.

In obtaining the solution formula (\ref{zsol}) of $z(s,\xi)$, we
avoid doing back transformation as described by \citet{lc}. Noticing
that if we want to specify the value of a matching prior at a
certain point, say $(\ipara^*, \npara^*)$, we can directly specify
$s$ as $\ipara^*$ and $\xi$ as $\npara^*$ in formula (\ref{zsol}),
and then the matching prior evaluated at $(\ipara^*,\npara^*)$ can
be achieved.

Without loss of generality, set the initial condition
$$\{\Psi(\xi),
\Lambda(\xi), Z(\xi)\}=(0,\xi,-1).$$
With $\Psi(\xi)=0$, we have
$\ipara=s$.  The equations (\ref{ode2}) and (\ref{ode3}) can be
simplified as
    \begin{align}
        \label{ode4}\frac{\der \npara}{\der s}=
            &\frac{b(s,\npara)}{a(s,\npara)},\ \ \
            \npara(s_0,\xi)=\Npara(\xi),\\
        \nonumber\label{ode5}\frac{\der z}{\der s}=
            &\frac{d(s,\npara)}{a(s,\npara)},
            \ \ \ z(s_0,\xi)=Z(\xi).
    \end{align}

We used R package \verb"odesolve"  by \citet{Rodesolve} to solve
equation (\ref{ode4}) and get a numerical expression of
$\npara(\cdot)$ in $s$.  The command \verb"lsoda()" in
\verb"odesolve" package is designed to solve initial value problems
for stiff or non-stiff systems of first order ordinary differential
equations. It provides an interface to the Fortran ordinary
differential equation solver of the same name, written by \citet{H}
and \citet{P}. For (\ref{zsol}), we did numerical integration using
Simpson's Rule and employed the R function \verb"sintegral()" in the
\verb"Bolstad" package by \citet{RBolstad}. Suppose $z$ will be
evaluated at $(\ipara^*, \npara^*)$. Noticing that
$\Npara(\xi)=\xi$, choose the start value as $\npara^*$ in solving
(\ref{ode4}), and then choose the upper integration limit as
$\ipara^*$ in (\ref{zsol}). The procedure is easy to perform if one
has an ordinary differential equation solver, even if not using the
solver provided by R package \verb"odesolve".

Based on the ordinary differential equation (\ref{ode1}),
$\ipara=s+\Psi(\xi)$, i.e. $s=\ipara-\Psi(\xi)$. So $s_0$ must be
chosen considering the range of $\ipara$. If we choose
$\Psi(\xi)=0$, then $\ipara=s$. For the first example in \S 5, the
parameter $\ipara$ is the ratio of two exponential means, and hence
$\ipara>0$. Therefore, $s_0$ should be chosen as any positive value.

In the above we choose the initial values as $\{\Psi(\xi),
\Lambda(\xi), Z(\xi)\}=(0,\xi,-1)$. Now we will show that the
numerical solving procedure is suitable to any initial values.

\noindent$\bullet$ Suppose the initial condition for the ordinary
differential equation (\ref{ode2}) is
$\npara(s_0,\xi)=\Lambda(\xi)$, for $\Lambda(\xi)$ an arbitrary
known function rather than $\Lambda(\xi)=\xi$ as above. The solution
formula of $z$ is the same as stated in (\ref{zsol}). When solving
(\ref{ode2}), the initial value should be chosen as
$\Lambda(\npara^*)$, no longer $\npara^*$, if $z$ is evaluated at
$(\ipara^*,\npara^*)$.

\noindent$\bullet$ If the initial condition of (\ref{ode1}) is
$\ipara(s_0,\xi)=\Psi(\xi)$, then the solution from the equation
(\ref{ode1}) is $\ipara=s+\Psi(\xi)$. Therefore, the equation
(\ref{ode2}) becomes,
    $$\frac{\der \npara}{\der s}
        =\frac{b\{s+\Psi(\xi),\npara\}}{a\{s+\Psi(\xi),\npara\}}.$$
Let $\tilde s = s+\Psi(\xi)$.  By simple change of variables,
(\ref{ode2}) becomes
    $$\frac{\der \npara}{\der \tilde s}
        =\frac{b(\tilde s,\npara)}{a(\tilde s,\npara)}.$$
Equation (\ref{ode3}) is
    $$\frac{\der z}{\der \tilde s}
        =\frac{d[\ipara\{\tilde s-\Psi(\xi),\xi\},
        \npara\{\tilde s-\Psi(\xi),\xi\}]}{a[\ipara\{\tilde s-\Psi(\xi),\xi\},
        \npara\{\tilde s-\Psi(\xi),\xi\}]}$$
with $z\{\tilde{s}_0-\Psi(\xi),\xi\}=Z(\xi)$, noticing that
$\tilde{s}_0=s_0+\Psi(\xi)$. Then the solution of $z$ is simply
given by the following formula,
    \begin{equation}\label{zsol11}
        z(\tilde s,\xi)=Z(\xi)+ \int^{\tilde s-\Psi(\xi)}_{s_0-\Psi(\xi)}
        \frac{d\{\ipara(s',\xi),\npara(s',\xi)\}}
                {a\{\ipara(s',\xi),\npara( s',\xi)\}}\der s'.
    \end{equation}
That is to say, the value of the prior on a certain point with the
initial condition $\ipara(s_0,\xi)=\Psi(\xi)$, is obtained by
translating the interval of integration when $\Psi(\xi)=0$ by
$\Psi(\xi)$.

\noindent$\bullet$ Suppose the initial condition for (\ref{ode3}) is
$z(s_0,\xi)=Z(\xi)$ and $Z(\cdot)$ is a known function.  This case
is even simpler to deal with.  One only needs to plug the value of
$Z(\xi)$ into (\ref{zsol}).

Therefore, the suggested numerical solving procedure is suitable to
any initial values.

In the above, both the parameter of interest and the nuisance
parameter are scalars.  With dimension 2, it is relatively easy to
understand the first order partial differential equation solving
procedure from the geometric point of view, since one can draw the
initial conditions and the solution surface in a 3-dimensional
space. In \citet{zhang}, the solving procedure was extended to
multiple nuisance parameters, while keeping the parameter of
interest as a scalar. The procedure of the higher dimension is
similar as the one of 2-dimensional model parameters. However, when
$d>2$, it can be computational intensive to implement the procedure.
Also, if there are no explicit expressions for the coefficients in
the original first order partial differential equation, numerical
implementation may be more difficult.

\vskip 0.8cm

\begin{center}
    \textsc{4. DiCiccio and Martin's approximations}
\end{center}

Likelihood ratio test is widely used in statistical inference.  The
signed root of the likelihood ratio statistic is
$R=\sgn(\hipara-\ipara_0)[2\{l(\hbpara)-l(\iparanull,\hbnpara_0)\}]^{1/2}$,
where $l(\bpara)$ is the log-likelihood function for the unknown
parameter vector $\bpara$ and $\hbnpara_0$ is shorthand for
$\hbpara_{\ipara_0}$, the constrained maximum likelihood estimator
of $\bpara$.  The standard normal approximation to the distribution
of $R$ typically has error of order $O(n^{-1/2})$, and $R$ can be
used to construct approximate confidence limits for $\ipara$ having
coverage error of order $O(n^{-1/2})$.

Using matching priors, \citet{DM} proposed tail probability
approximations of order $O(n^{-1})$.  The approximations are
saddlepoint approximations that involve Bayesian method. The
approximations of \citet{DM} can be expressed in the \citet{bn80}
format
    \begin{equation}\label{bnformat}
        \Phi\{R+R^{-1}\log(T/R)\},
    \end{equation}
and the \citet{lr80} format
    \begin{equation}\label{lrformat}
        \Phi(R)+\phi(R)(R^{-1}-T^{-1}),
    \end{equation}
where $\Phi$ is the standard normal distribution function, and $T$
is defined as
    \begin{align}\label{T}
        T=l_{\ipara}(\ipara_0,\hbnpara_0)
        \frac{|-l_{\bnpara\bnpara}(\ipara_0,\hbnpara_0)|^{1/2}\pi(\hbpara)}
        {|-l_{\bpara\bpara}(\hbpara)|^{1/2}\pi(\ipara_0,\hbnpara_0)}.
    \end{align}
Here $l_{\ipara}(\bpara)=\partial l(\bpara)/\partial \ipara$,
$l_{\bpara \bpara}$ is the matrix of second-order partial
derivatives of $l(\bpara)$ taken with respect to $\bpara$;
$l_{\bnpara \bnpara} (\bpara)$ is the submatrix of $l_{\bpara
\bpara}(\bpara)$ corresponding to $\bnpara$; and $\pi(\bpara)$ is a
matching prior density for $\bpara=(\ipara,\bnpara)$ which satisfies
equation (\ref{prior}). Then the resulting approximation is
$\Prob(\ipara\geq\ipara_0|X)\doteq\Phi\{R+R^{-1}\log(T/R)\},$ or,
$\Prob(\ipara\geq\ipara_0|X)\doteq\Phi(R)+\phi(R)(R^{-1}-T^{-1}).$
Both of them have relative error of order $O(n^{-1})$.  Approximate
confidence limits for $\ipara$ can be constructed using either of
(\ref{bnformat}) or (\ref{lrformat}). These confidence limits have
coverage errors of order $O(n^{-1})$. To relative error of the order
$O_p(n^{-1})$, the variable $T$ is parameterization invariant under
transformations $\bpara \mapsto \{\psi,\tau(\bpara)\}$.

The approximations of \citet{DM} show their advantages in less
computational effort compared to the Metropolis-Hastings procedure
used by \citet{lc}. To calculate $T$ in (\ref{T}), the matching
prior requires to be evaluated at two points,
$(\ipara_0,\hbnpara_{\ipara_0})$ and $\hbpara$.  The initial curve
can be chosen passing through $(\ipara_0,\hbnpara_{\ipara_0})$; that
is to say, only the solution on one point $\hbpara$ needs to be
determined.

\vskip 0.8cm

\begin{center}
    \textsc{5. Examples}
\end{center}

\begin{center}
    {5.1. \textit{Ratio of two exponential means}}
\end{center}

Let $X$ and $Y$ be exponential random variables with means $\mu$ and
$\nu$ respectively; the ratio of the means, $\nu/\mu$, is the
parameter of interest.  The parameter transformation $\left(\mu\to
\lambda\ipara^{-\frac 12},\ \nu\to \lambda\ipara^{\frac 12}\right)$
makes the two new parameters $\ipara$ and $\lambda$ orthogonal. Then
$X$ and $Y$ have expectations $\lambda\ipara^{-\frac 12}$ and
$\lambda\ipara^{\frac 12}$, respectively.

Suppose we have $n$ independent replications of $(X, Y)$.  Denote
$\bpara=(\ipara,\lambda)$. We can obtain the log-likelihood function
as
$l(\bpara)=-n\left\{(\ipara\bar{x}+\bar{y})/(\lambda\surd{\ipara})+2\log
\lambda\right\}.$

Both approximations of the Barndorff-Nielson format (\ref{bnformat})
and the Lugannani and Rice format (\ref{lrformat}) are considered.
Based on these approximations, $p$-values can be calculated.
Approximations based on different prior density functions mentioned
previously may be used to generate an approximate one-sided
$p$-value by approximating $\Prob (R\ge r)$, for $r$ the observed
value of $R$. Approximate two-sided $p$-values may be calculated by
approximating $2\min \{\Prob(R\ge r), \Prob(R < r)\}$. One and
two-sided hypotheses tests of size $\alpha$ may be constructed by
rejecting the null hypothesis when the $p$-value is less than
$\alpha$.  Table \ref{exptypeI} reports type I error probabilities
of the 1,000,000 rounds of simulation with $n=10$.

In this example, the parameters $\ipara$ and $\npara$ are
orthogonal. Using the simplified partial differential equation
(\ref{othprior}), $\pi(\ipara,\lambda)=1/\ipara$ is an explicit
solution.  Also $\pi(\ipara,\lambda)=1/(\ipara \lambda)$ is another
explicit solution. Numerical solutions were also calculated. One of
the initial condition is $\{\Psi(\xi),\Lambda(\xi),Z(\xi)\}=(0, \xi,
-1)$. The resulting matching prior corresponds to the the analytic
solution $1/\ipara$. Another numerically solved matching prior is
based on the initial condition $(0, \xi, -\log\xi)$, which
corresponds to the the analytic solution $1/(\ipara\npara)$. From
Table \ref{exptypeI}, one can see that the numerical and analytic
solutions give almost the same simulation results, which confirmed
the validity of our numerical solution process.

Approximations (\ref{bnformat}) and (\ref{lrformat}) have a
removable singularity at $R=0$. Consequently, these and similar
formulae require care when evaluating near $R=0$. In these cases,
for all but the most extreme conditioning events, the resulting
conditional $p$-value is large enough as to not imply rejection of
the null hypothesis, and so these simulated data sets are treated as
not implying rejection of the null hypothesis.

\begin{center}
    {5.2. \textit{Logistic regression}}
\end{center}

We consider a logistic regression model with a binary response $Y$
and only one explanatory variable $X$. Let $\para_1$ denote the
unknown intercept and $\para_2$ denote the unknown effect of the
explanatory variable. Suppose $\para_2$ is the parameter of interest
and $\para_1$ is the nuisance parameter.  We will solve matching
priors and apply DiCiccio and Martin's approximations to do
inference about $\para_2$. \citet{lc} considered a similar example.

Let $Y_i$ be the response variable taking binary values with success
probability as $p_i$, and $X_i$ be the explanatory variable
following uniform distribution $U(0,1)$. Suppose there are $n$
independent replications of $(X_i, Y_i)$.  Fit the model
$\log\{p_i/(1-p_i)\}=v_i'\bpara=\para_1+\para_2 x_i$, where
$v_i=(1,x_i)'$ and $\para_2$ is the parameter of interest. Inverting
the equation, we have $p_i=(1+e^{-v_i'\bpara})^{-1}$.  We can obtain
the log-likelihood function as $l(\bpara;x)=\sum_{i=1}^n y_i\log\{
        p_i/(1-p_i)\}+\sum_{i=1}^n \log(1-p_i).$
The first derivative of the log=likelihood function is $V'(y-p)$,
where $V$ is the design matrix with $v_i'$ in row $i$. The second
derivative of log-likelihood function is $-V'WV$, where $W$ is a
diagonal matrix with diagonal elements $p_i(1-p_i), i=1,\cdots,n$.

Using sample size $n=30$, we generate data satisfying the logistic
regression model with $\para_1=-1$, $\para_2=0.5$, and the
explanatory variable $X$ following uniform distribution $U(0,1)$. In
this case, generally the parameters $\para_1$ and $\para_2$ are not
orthogonal. We use the numerical procedure described in \S 3 and
study performances of different initial conditions.  Table
\ref{logistictypeI} contains type I error probabilities for both
one-sided and two-sided tests for approximations of both
Barndorff-Nielson format and Lugannani and Rice format, based on
10,000 rounds of simulation.

As we mentioned previously, approximations (\ref{bnformat}) and
(\ref{lrformat}) have a removable singularity at $R=0$.  We deal
with this singularity the same way as in \S 5.1.

In the following, we give some instructions on how to change the
initial condition and how to choose favorable initial conditions.
Initial condition $(0,\xi,-1)$ gives type I error probabilities
larger than the nominal level 0.05; that is to say, it has the
tendency to underestimate tail probabilities and reject the null
hypothesis. We want to choose initial conditions to obtain a test
whose type I error rate is closer to the nominal level.  We adjust
the initial condition when solving the partial differential equation
(\ref{prior}), and use the Barndorff-Nielson format of the
approximation. The quantity $T$ in (\ref{T}) is the only part in the
approximation that relates to matching priors. For a one-sided test,
when the probability is small and close to 0, $R$ and $T$ are
negative. Making $\Phi\{R+R^{-1}\log(T/R)\}$ larger is equivalent to
making $T$ bigger. Also one may notice that $Z(\xi)$ is used only in
equation (\ref{zsol}).  Suppose the initial condition is
$\{\Psi(\xi), \Lambda(\xi), Z(\xi)\}$.  Keep the first two
components of the initial condition, $\Psi(\xi)$ and $\Lambda(\xi)$,
unchanged, and only modify the third term, $Z(\xi)$. By doing so,
the integral part in equation (\ref{zsol}) is kept unchanged and $z$
varies only with $Z(\xi)$. By changing $Z(\xi)$, we want to adjust
$T$ to be bigger. Because $T$ is negative when reject a hypothesis,
and matching priors appear in $T$ as a ratio, one can construct a
$Z(\cdot)$ such that the ratio,
$\exp\{Z(\hat{\ipara},\hat{\npara})\} /
\exp\{Z(\ipara_0,\hat{\npara}_0)\}$, will be smaller than 1; recall
that 1 is the value of the ratio when $Z(\xi)=-1$. Based on the
above arguments, $Z(\cdot)$ function is constructed as
$Z(\xi)=-\log\{(\xi+1)^q+1\}$, where $q$ is a tuning parameter and
leads $Z(\cdot)$ to an even function.  As an even function, $Z(\xi)$
achieves its maximum value at $-1$, where $-1$ is the true value for
the nuisance parameter when data were simulated. We have constructed
priors using knowledge of the true value of the nuisance parameter.
Of course, in practice this knowledge is unavailable. One might
instead use an estimator of the nuisance parameter in place of the
true value.

When $Z(\xi)$ increases quickly, such as $q=2$ in table
\ref{logistictypeI}, the type I error probability deviates far away
from the nominal level in the other direction.  If a more slowly
increasing functions is used, the performance of type I error may be
better.

Unfortunately, with some choices of initial conditions, such as the
last three listed in table \ref{logistictypeI}, the Lugannani and
Rice format approximation may fall outside the range of 0 and 1 in
some cases.  For example, the initial condition of
$[0,\xi,-\log\{(\xi+1)^{2}+1\}]$ yielded 5 such probabilities out of
10,000 data sets.  We convert those values to 0 or 1 by
$\min\{\max(p,0),1\}$, where $p$ is the $p$-value that is outside 0
and 1.

For the parameter of interest $\para_2$, we calculate credible
intervals using DiCiccio and Martin's approximation in
Barndorff-Nielson format. With initial condition $(0,\xi,-1)$, out
of 1,000 generated data sets, there are 938 credible intervals
covered the true value $0.5$.  With initial condition
$[0,\xi,-\log\{(\xi+1)^{2/5}+1\}]$, for the parameter of interest
$\para_2$, there are 954 credible intervals covered the true value
$0.5$.

We apply the above procedure to a real data set from \citet[Table
1.1]{HL}. The response variable is coronary heart disease indicator,
$y$, and the explanatory variable is age, $x$. One hundred subjects
were included in the study; i.e. $n=100$.  We fit the logistic
regression model following the same definition as above, with
$\para_1$ defined for the unknown intercept and $\para_2$ for the
effect of age on heart disease status.  Using initial condition
$(0,\xi,-1)$ and Barndorff-Nielson format approximation, a two-sided
testing $p$-values is $5.532326\times 10^{-8}$, and five and
ninety-five posterior percentiles are of 0.07 and 0.15 respectively.

\vskip 0.8cm
\begin{center}
    \textsc{6. Conclusion}
\end{center}
Matching priors were first proposed by \citet{WP} and \citet{Pe}. In
the general parametrization, if the parameter of interest and the
nuisance parameters are not orthogonal, solving the prior from a
first order partial differential equation is nontrivial. This paper
presents a practical way to solve for the matching priors and the
procedure can be suitable to all kinds of initial conditions.
Matching priors can be used with the approximations of \citet{DM}.
By choosing differential initial conditions one is able to improve
the performances of DiCiccio and Martin's approximations.

\bibliography{bib}

\newpage

\begin{table}[!ht]
\centering\caption{Ratio of two exponential means: type I error
probability}\label{exptypeI}
\begin{tabular} {|l|l|l|l|l|}
\hline {} & \multicolumn {2}{|c|} {BN Format} &
            \multicolumn {2}{|c|} {LR Format}\\
\hline {Tests} & 1-sided& 2-sided  & 1-sided & 2-sided\\
\hline Likelihood ratio test
                            &0.0520  &0.0526  &0.0520  &0.0526\\
\hline I.C. $(0,\xi,-1)$
                            &0.0456  &0.0441  &0.0456  &0.0441\\
\hline Analytic solution: $1/\ipara$
                            &0.0456  &0.0441 &0.0456  &0.0441\\
\hline I.C. $(0,\xi,-\log\xi)$
                            &0.0499  &0.0498 &0.0499 &0.0498\\
\hline Analytic solution: $1/(\ipara\npara)$
                            &0.0499  &0.0498 &0.0499 &0.0498\\
\hline \multicolumn{5}{l}{\footnotesize{$^*$I.C. stands for
initial condition.}}\\
\multicolumn{5}{l}{\footnotesize{$^{\dag}$Results are based on
1,000,000 rounds of simulation with $n=10$.}}\\
\multicolumn{5}{l}{\footnotesize{$^{\ddag}$Tests are of nominal type I error 0.05.}}\\
\end{tabular}
\end{table}

\newpage
\begin{table}[!ht]
\centering\caption{Logistic regression: type I error
probability}\label{logistictypeI}
\begin{tabular} {|l|l|l|l|l|}
\hline {} & \multicolumn {2}{|c|} {BN Format}
        & \multicolumn {2}{|c|} {LR Format}\\
\hline {Test} & 1-sided& 2-sided &  1-sided & 2-sided\\
\hline Likelihood ratio test
                                &0.054 &0.060 &0.054 &0.060\\
\hline I.C. $(0,\xi,-1)$
                                &0.052 &0.057 &0.052 &0.057\\
\hline I.C. $[0,\xi,-\log\{(\xi+1)^{2}+1\}]$
                                &0.028 &0.019 &0.031 &0.020\\
\hline I.C. $[0,\xi,-\log\{(\xi+1)^{2/5}+1\}]$
                                &0.041 &0.041 &0.044 &0.046\\
\hline I.C. $[0,\xi,-\log\{(\xi+1)^{2/11}+1\}]$
                                &0.045 &0.048 &0.046 &0.050\\
\hline \multicolumn{5}{l}{\footnotesize{$^*$I.C. stands for initial condition.}}\\
\multicolumn{5}{l}{\footnotesize{$^{\dag}$Results are based on
10,000 rounds of simulation with $n=30$.}}\\
\multicolumn{5}{l}{\footnotesize{$^{\ddag}$Tests are of nominal type I error 0.05.}}\\
\end{tabular}
\end{table}

\end{document}